\documentclass[preprint,12pt]{elsarticle}

\usepackage{amssymb}
\usepackage{amsmath}
\usepackage{color}
\usepackage{ulem} %

\begin{document}

\begin{frontmatter}



\title{Generalized Mpemba effect in diffusion-controlled spin-dependent delayed fluorescence}


\author[aist]{Kazuhiko Seki}

\address[aist]{GZR, National Institute of Advanced Industrial Science and Technology (AIST), Onogawa 16-1 AIST West, Tsukuba, Ibaraki 305-8569, Japan}
\begin{abstract}
The Mpemba effect is commonly associated with anomalous thermal relaxation, in which a system prepared at a higher temperature reaches equilibrium faster than one prepared at a lower temperature. In modern formulations, however, its defining feature is broader: a state initially farther from equilibrium can relax faster than a closer one, or relaxation trajectories can cross. 
Here, we show that magnetic-field-dependent delayed fluorescence in triplet-fusion systems realizes a generalized Mpemba effect driven by external control of spin-selective kinetics. Using a diffusion-controlled geminate-recombination extension of the Johnson--Merrifield model, we demonstrate that delayed-fluorescence trajectories obtained at different magnetic fields cross when geminate fusion is effective. We further derive a compact kinetic criterion for such crossing in terms of the competition between intermediate-time decay rates and long-time power-law amplitudes. This criterion captures the interplay between effective fast and slow relaxation contributions. Within this kinetic framework, the final state is independent of the external control parameter and is defined as the fully relaxed state in which no separated triplet population remains. The generalized Mpemba effect therefore arises from the redistribution of transient relaxation pathways associated with geminate recombination, whereas bulk diffusion-controlled processes contribute only to the final relaxation.

These results link anomalous relaxation in spin-dependent photokinetics to modern formulations of the generalized Mpemba effect and show how trajectory crossing can emerge in systems governed by diffusion-controlled geminate recombination and non-Markovian kinetics.
\end{abstract}

\begin{graphicalabstract}
\includegraphics[width=0.5\columnwidth]{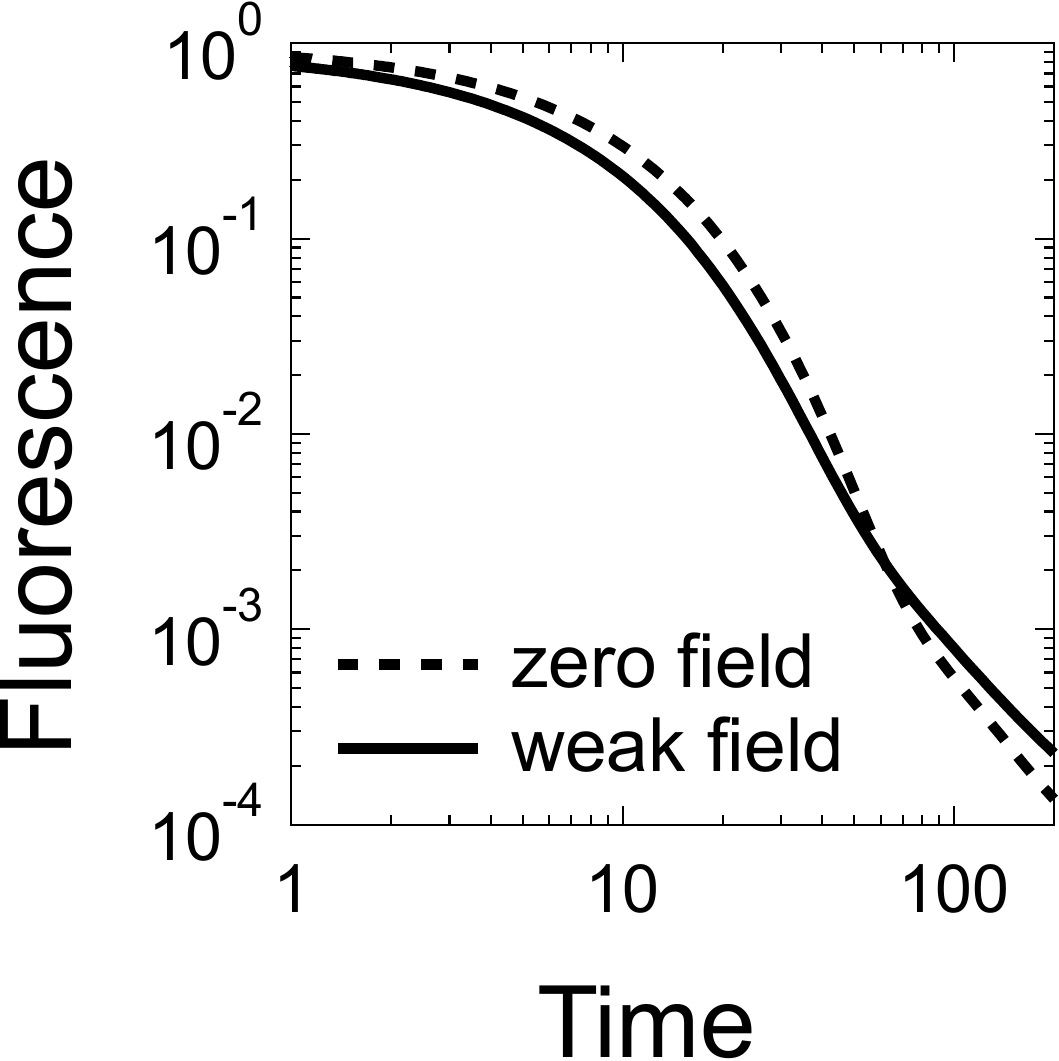}
\end{graphicalabstract}

\begin{highlights}
\item Magnetic-field-dependent delayed fluorescence shows trajectory crossing.
\item Geminate recombination produces power-law relaxation.
\item Fast decay competes with a slow diffusion-controlled contribution.
\item Spin-selective kinetics redistributes relaxation pathways.
\end{highlights}

\begin{keyword}
Mpemba effect \sep magnetic field effect \sep anomalous relaxation \sep diffusion \sep geminate recombination \sep triplet fusion



\end{keyword}

\end{frontmatter}



\section{Introduction}

The Mpemba effect is traditionally introduced as an anomalous thermal relaxation process in which a system prepared at a higher temperature reaches equilibrium faster than an identical system prepared at a lower temperature \cite{mpemba1969,lu2017mpemba}. 
In modern formulations, however, the essence of the phenomenon is not temperature itself but the overtaking of relaxation trajectories  (Mpemba crossings): a state initially farther from equilibrium can relax faster than a closer one.  \cite{TEZA_26,Wang_26}

 In the present work, we study trajectory crossing of an observable during relaxation from a common initial state to a common final state. The energy levels of the intermediate states can be closer to that of the final state, yet their relaxation can nevertheless be slower, giving rise to trajectory crossing. Although it is the intermediate states, rather than the initial state, that are modulated here, we refer to this crossing as the generalized Mpemba effect, while noting that it differs from the definition in which a closer initial state relaxes more slowly.

This trajectory-based viewpoint has been established in recent theoretical studies, where generalized Mpemba effects are characterized by contributions from relaxation modes other than the slowest mode \cite{hayakawa2023}. 
More complex behavior, including multiple crossings, can arise in systems with multi-mode or oscillatory relaxation dynamics \cite{hayakawa2024multiple}; analytically tractable models have also been developed, for example in two-dimensional bistable potentials, where crossing is governed by relaxation-mode amplitudes \cite{hayakawa2026}.

The generalized Mpemba effect has also been reported to be controlled by the projection of the initial distribution onto the system's relaxation modes: in single-particle systems described by Langevin or master equations in confining potentials, with and without activity \cite{brownian_singlewell2023,langevin_metastability2023,active_markov2025,active_brownian_notrap2025}; in many-particle systems such as driven granular gases \cite{granular_maxwell2020,granular_anisotropic2021}; and in a trapped-ion qubit realization, where quantum coherence produces a non-monotonic overlap with the slow relaxation mode \cite{trapped_ion_inverse2024}. A recent comprehensive review summarizes this rapidly growing literature \cite{TEZA_26}.

Crossing of relaxation trajectories has also been observed in the photokinetics of singlet fission and subsequent triplet-pair fusion \cite{burdett2013,piland2013}. 
Singlet fission occurs when the energy of the lowest singlet excited state is comparable to, or greater than, twice the triplet-state energy, while the reverse process is known as triplet (pair) fusion. 
Magnetic-field-dependent fluorescence decay measurements following pulsed excitation to the singlet excited state have shown that fluorescence decay curves measured at different magnetic fields can cross in time, indicating intersecting relaxation trajectories toward the singlet ground state, as reported experimentally by Burdett, Piland, and Bardeen \cite{burdett2013,piland2013}.

A microscopic description of this behavior has been developed using diffusion-controlled geminate-pair recombination models of triplet-fusion kinetics \cite{seki2018,shushin2018,seki2021}. These models show that trajectory crossing arises from diffusion-controlled geminate recombination coupled to spin-selective reaction pathways.

Within this framework, crossing originates from a redistribution of trajectory weights among spin-selective pathways, which determines the magnitude of the long-time power-law decay generated by diffusion-controlled geminate recombination. Although the initial and final singlet states are not directly affected by the magnetic field, the triplet-pair spin states are modified by the applied field through the Zeeman interaction. In particular, the magnetic field changes spin mixing and thereby alters the relative contributions of kinetic pathways that regenerate singlet excited states from geminate triplet pairs.

As a result, an initial fast decay associated with singlet fission is followed by delayed regeneration of singlet excited states through diffusion-controlled geminate recombination and spin-selective conversion, leading to crossing of the relaxation trajectories. Such crossing has been shown to occur when the triplet-fusion rate constant governing singlet excited-state repopulation is sufficiently large compared with the dissociation rate constant of the triplet pair \cite{seki2018}.

Motivated by this connection, we revisit magnetic-field effects in triplet fusion systems. 
Previous studies have demonstrated trajectory crossing in the high-field regime, where the delayed fluorescence ultimately decays faster under the magnetic field. 
Here, we show that, in the weak-field regime, the fluorescence initially decays faster than in the zero-field case, subsequently crosses it, and eventually relaxes more slowly toward the common final state, in which triplets are converted back to the singlet ground state.  
Magnetic fields modify the transient spin-selective pathways, 
whereas the fully relaxed population is assumed to reside in the singlet ground state irrespective of field strength.

\label{remark:initial_condition} 
The initial condition is the singlet excited-state population created by pulsed excitation and is magnetic-field independent, since the singlet state itself is not influenced by the magnetic field. 
Singlet fission occurs with a time constant of approximately $1$--$100$ ps and has no magnetic-field dependence. 
Subsequent spin-mixing time for closely spaced triplet pairs may be on the order of picoseconds to tens of picoseconds, depending on the exchange interaction, dipolar coupling, molecular orientation, and magnetic field. 
On the other hand, triplet fusion is a slower process and gives magnetic-field dependence to diffusion-controlled delayed fluorescence after about $10$ ns following pulsed excitation. \cite{johnson1970,merrifield1971,steiner1989}
In the long-time limit, all excitons return to magnetic-field independent singlet ground states. 

Because singlet fission is fast and magnetic-field independent while the subsequent fast spin mixing modifies the triplet-fusion kinetics, the magnetic-field dependence of the photoluminescence originates from the slower triplet-fusion process. Although the singlet state itself is not directly affected by the magnetic field, the applied field lowers the energy of the lowest triplet sublevel through Zeeman splitting;  
nevertheless, recovery to the common singlet ground state is slower under a weak magnetic field, giving rise to the trajectory crossing that we identify with the generalized Mpemba effect. 
We further show that the direction of this crossing reverses between the weak-field and high-field regimes: the high-field crossing, in which the delayed fluorescence ultimately decays faster under the magnetic field, can likewise be regarded as a realization of the generalized Mpemba effect in its broader trajectory-crossing-based definition. 
Although the field-dependent spin mixing that follows singlet fission could, in principle, be regarded as generating an effectively field-dependent initial condition for the subsequent triplet-fusion and diffusion-controlled recombination dynamics that exhibit the crossing, this post-mixing triplet-pair state cannot be prepared or controlled directly by experiment: singlet fission from the common, magnetic-field-independent excited singlet population $S(0)$ is the only experimental means of generating it. We therefore adopt $S(0)$ as the initial condition throughout this work. This experimental procedure for setting the effectively field-dependent initial condition of the triplet-fusion process -- singlet fission followed by fast spin mixing -- is guaranteed to be valid only insofar as the time-scale separation between fission-and-mixing and fusion-and-diffusion-controlled recombination holds; we verify this assumption directly in \ref{app:reverse_ic} by comparing the dynamics computed from $S(0)$ against dynamics computed by imposing each field's post-mixing triplet-pair distribution directly as the initial condition.

Comparison of the weak-field and zero-field cases thus constitutes a concrete realization of the generalized Mpemba effect in a spin-dependent kinetic system. 
Magnetic-field-dependent delayed fluorescence therefore provides realizations of 
the same underlying kinetic scheme, distinguished only by the external control parameter $B$ (magnetic field strength), while the observable fluorescence, proportional to the singlet population, relaxes toward a common final state, here defined as the state in which no separated triplet population remains.

To formulate the generalized Mpemba effect as trajectory crossing toward a common final state, we define the final state as a fully relaxed state in which no separated triplet population remains. 
Within this framework, we focus on diffusion-controlled geminate recombination as the process that determines the observable dynamics, and treat bulk diffusion-controlled recombination, in which recombination occurs independently of the original pairing, as contributing only to the final relaxation without inducing trajectory crossing. 
Under this assumption, the fully relaxed state is independent of the external control parameter, and the generalized Mpemba effect can be attributed solely to the redistribution of transient relaxation pathways rather than to differences in the final state.

Building on diffusion-controlled geminate recombination extensions of the Johnson--Merrifield framework, we show that the observed kinetic trajectory crossing can be naturally interpreted 
as a generalized Mpemba effect driven by magnetic-field control of spin-selective fusion kinetics. 
We further derive a compact kinetic criterion that determines when such trajectory crossing occurs, thereby establishing a direct connection between spin-dependent photokinetics and general theories of anomalous relaxation.

The new contribution of the present work is not the observation of kinetic-curve crossing itself, but its formulation as a generalized Mpemba effect governed by competition between an effective transient decay and a diffusion-controlled power-law tail, together with the resulting compact crossing criterion. In conventional theories of the generalized Mpemba effect, relaxation dynamics are typically described in terms of a finite set of modes, as in Markovian or finite-state systems \cite{lu2017mpemba,hayakawa2023}. In contrast, in the present photokinetic system, triplet-pair regeneration proceeds via diffusion-controlled geminate recombination in an effectively infinite space, giving rise to long-time power-law decay. The repeated dissociation and regeneration of triplet pairs therefore introduce non-Markovian kinetics, which modify the relative weights of relaxation pathways and naturally lead to trajectory crossing.
 Concretely, the non-Markovian character referred to throughout this manuscript originates from the diffusion-controlled regeneration of associated triplet pairs: after dissociation, the probability that a separated triplet pair re-encounters and regenerates a correlated pair depends on the entire history of the diffusive motion, rather than on the instantaneous population alone, and this re-encounter probability decays as a power law at long times in an effectively infinite space. This is in contrast to the Markovian limit of the Johnson--Merrifield model, in which the diffusion-controlled regeneration process is replaced by an effective first-order rate constant and the possibility of geminate re-encounter after dissociation is neglected. In Ref.~\cite{nonMarkovian2025}, the source of non-Markovianity is the eliminated degrees of freedom of a quantum bath, traced out of the joint system-bath unitary dynamics. In the present work, the eliminated degree of freedom is instead the spatial coordinate of the dissociated triplet pair: the joint (reaction $+$ diffusion) process is Markovian, but projecting onto the population variables $S(t)$, $C^{(j)}(t)$, and $f^{(j)}(t)$ alone leaves a memory kernel, $R(t)$ in Eq.~(\ref{eq:g_1}), that encodes the diffusive return probability. It is this classical, spatial-diffusion projection, rather than the projection of a quantum bath, that we invoke when referring to non-Markovian kinetics in the present system.

Given that the Mpemba effect is highly sensitive to experimental conditions and not universally observed, and that its realization is often difficult to reproduce reliably \cite{Bechhoefer2021,Ares2025}, establishing its signature in systems with well-defined microscopic dynamics, such as spin-dependent relaxation in triplet-pair systems, is of particular importance for elucidating its general kinetic origin.

\section{Theory and mode-competition viewpoint}

\begin{figure}
\centerline{
\includegraphics[width=0.6\columnwidth]{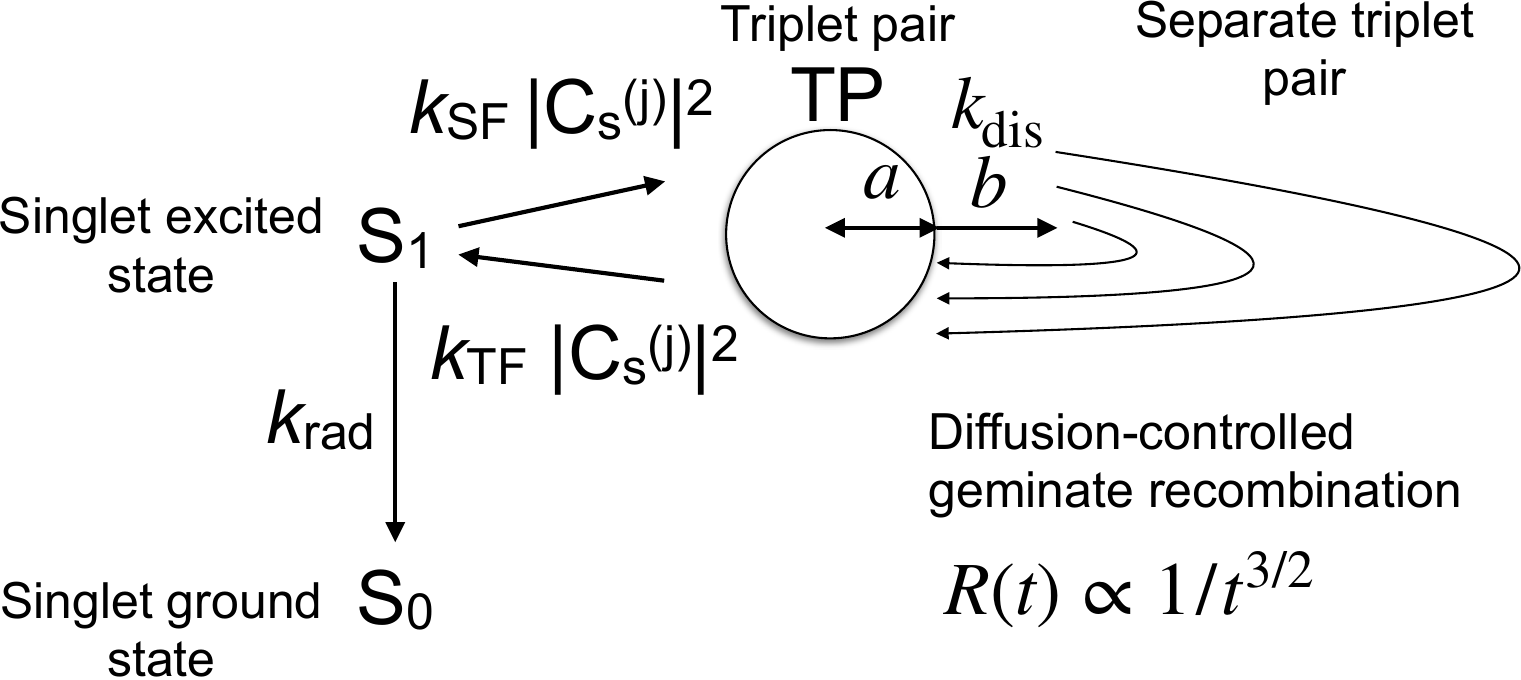}
}
\caption{Kinetic model describing singlet fission and triplet fusion, 
in which a triplet pair dissociates into a separated triplet pair 
and is subsequently regenerated by diffusion-controlled geminate recombination. 
Fluorescence with the rate constant $k_{\rm rad}$ exhibits a delayed component owing to the power-law decay 
$R(t) \propto t^{-3/2}$ of the first-passage-time distribution for the separated geminate pair 
to return from the initial separation $a+b$ to the encounter distance $a$.
} 
\label{fig:scheme}
\end{figure}

Here, we review the delayed-fluorescence kinetics due to triplet fusion in Ref.~\cite{seki2018} and reformulate the results from the viewpoint of the generalized Mpemba effect. 
As shown in Fig.~\ref{fig:scheme}, we consider a spin-selective kinetic description of singlet fission (SF) and triplet (pair) fusion (TF), in which the initially excited singlet population $S(t)$ is coupled to an associated triplet-pair population $C^{(j)}(t)$ and a separated (diffusing) triplet-pair population $f^{(j)}(t)$ for each spin state $j$.

In the Johnson--Merrifield model, 
coherent spin evolution within the triplet-pair manifold is not followed explicitly. 
Instead, spin mixing is assumed to be sufficiently fast, or effectively averaged, on the time scale of the kinetic processes considered. 
Off-diagonal elements of the spin Hamiltonian are neglected. 
The total associated and separated triplet-pair densities are $C(t)=\sum_{j=1}^9 C^{(j)}(t)$ and $f(t)=\sum_{j=1}^9 f^{(j)}(t)$, respectively. 
Spin selectivity enters through the singlet character $|C_{\rm s}^{(j)}|^2$ of each triplet-pair spin state, with $\sum_{j=1}^9 |C_{\rm s}^{(j)}|^2=1$. 
The diffusion-controlled extension of the Johnson--Merrifield model for geminate pairs reads \cite{seki2018}
\begin{eqnarray}
\frac{\partial}{\partial t} S(t)
&=& -\left(k_{\rm SF}+k_{\rm rad}\right)S(t)
+ \sum_{j=1}^9 k_{\rm TF} |C_{\rm s}^{(j)}|^2 \, C^{(j)}(t),
\label{eq:Sd}\\
\frac{\partial}{\partial t} C^{(j)}(t)
&=& k_{\rm SF} |C_{\rm s}^{(j)}|^2 S(t)
-\left(k_{\rm TF} |C_{\rm s}^{(j)}|^2+k_{\rm dis}\right) C^{(j)}(t)
+ k_g^{(j)}(t),
\label{eq:Cd}\\
\frac{\partial}{\partial t} f^{(j)}(t)
&=& k_{\rm dis} C^{(j)}(t) - k_g^{(j)}(t),
\label{eq:Fd}
\end{eqnarray}
where $k_{\rm rad}$, $k_{\rm SF}$, $k_{\rm TF}$, and $k_{\rm dis}$ denote the radiative decay, singlet fission, triplet fusion, and dissociation rate constants, respectively. 
We analyze the relaxation dynamics of $S(t)$ starting from the initial population $S(0)$.

The key non-Markovian ingredient is the diffusion-controlled regeneration term $k_g^{(j)}(t)$, written as a convolution with the regeneration kernel $R(t)$ \cite{seki2018}:
\begin{eqnarray}
k_g^{(j)}(t) \equiv \int_0^t dt_1 \, R(t-t_1)\, k_{\rm dis}\, C^{(j)}(t_1).
\label{eq:g_1}
\end{eqnarray}
Here $R(t)$ is the first-passage-time distribution for the separated geminate pair to return from the initial separation $a+b$ to the encounter distance $a$. 
Its Laplace transform is known in $d$ dimensions as \cite{redner_2001,seki2018}
\begin{eqnarray}
\hat{R}(s)=\left(\frac{a+b}{a}\right)^{1-d/2}
\frac{{\rm K}_{1-d/2}\!\left[(a+b)\sqrt{s/D}\right]}
{{\rm K}_{1-d/2}\!\left[a\sqrt{s/D}\right]},
\label{eq:FPTD}
\end{eqnarray}
where $D$ is the relative diffusion coefficient and ${\rm K}_\nu$ is the modified Bessel function of the second kind \cite{Abramowitz}.

Eliminating $C^{(j)}(t)$ and $f^{(j)}(t)$ in the Laplace domain yields the closed expression \cite{seki2018}
\begin{eqnarray}
\frac{\hat{S}(s)}{S(0)}
=
\left[
s+k_{\rm rad}+k_{\rm SF}
-\sum_{j=1}^9
\frac{k_{\rm SF}k_{\rm TF} |C_{\rm s}^{(j)}|^{4}}
{s+k_{\rm TF} |C_{\rm s}^{(j)}|^2+k_{\rm dis}\left(1-\hat{R}(s)\right)}
\right]^{-1}.
\label{eq:LaplaceS}
\end{eqnarray}

\subsection{Mode competition viewpoint: identifying ``fast'' and ``slow'' contributions}
\label{sec:modecomp}

Equation~(\ref{eq:LaplaceS}) shows that the relaxation contains two qualitatively distinct kinetic contributions: a discrete exponential component associated with a pole of Eq.~(\ref{eq:LaplaceS}) in the absence of the regeneration kernel, $\hat{R}(s)=0$, and a non-analytic contribution originating from the square-root dependence on $s$ through $\hat{R}(s)$. The latter effectively represents a continuum of relaxation processes rather than a single discrete mode.

\paragraph*{Field-controlled redistribution of weights and trajectory crossing}
The magnetic field modifies the singlet characters $C_{\rm s}^{(j)}(B)$ through spin mixing, thereby altering both the short-time (effective fast) contribution and the long-time regime via the amplitude of the diffusion-controlled power-law decay, without affecting the common final state, reflecting the conversion to the singlet ground state.
As a result, a trajectory that decays more slowly at early times can nevertheless exhibit a smaller long-time contribution, eventually leading to trajectory crossing. 
Conversely, numerical analysis indicates that a trajectory that decays more rapidly at early times can exhibit a larger long-time contribution.

In the modern formulation of the generalized Mpemba effect, such behavior corresponds to trajectory crossing toward a common final state. 
Thus, the crossing of fluorescence relaxation trajectories can be interpreted as a realization of the generalized Mpemba effect in a spin-dependent photokinetic system. 
This behavior reflects the redistribution of relaxation pathways enabled by the presence of an effective slow diffusion-controlled channel associated with geminate recombination.
\section{Generalized Mpemba criterion}
\label{sec:criterion}

In this section, we formulate a generalized Mpemba criterion for magnetic-field-dependent fluorescence decay. 
We consider the singlet population $S_B(t)$ (proportional to the delayed fluorescence intensity) relaxing toward a common final state under different values of the external control parameter $B$, which denotes the magnetic field strength. 
The essential point is that the relaxation dynamics consist of two qualitatively distinct contributions: an effective short-time (fast) component characterized by a finite decay rate constant, and a long-time (slow) geminate pair diffusion-induced contribution characterized by a power-law tail.

\subsection{Short-time region: effective transient decay rate constant}
\label{subsec:criterion_fast}

At sufficiently short times (before the geminate diffusion tail dominates), the decay can be described by an effective transient rate constant. 
From the expansion of the Laplace-domain solution in the large-$s$ limit, one obtains the initial rate 
 of singlet fission leading to a triplet pair
\begin{eqnarray}
k_{\rm ini}=k_{\rm rad}+k_{\rm SF},
\label{eq:kini}
\end{eqnarray}
which is independent of the singlet character $C_{\rm s}^{(j)}(B)$. 

By taking into account further dissociation of the triplet pair occuring  in competition to triplet fusion, 
the decay  at  early times is 
 characterized by the next-order transient rate constant \cite{seki2018}
\begin{eqnarray}
k_{\rm tr}(B)=k_{\rm rad}+k_{\rm SF}+k_{\rm fiss}^{(t)}(B),
\label{eq:ktrB}
\end{eqnarray}
where
\begin{eqnarray}
k_{\rm fiss}^{(t)}(B)=\sum_{j=1}^9
\frac{k_{\rm SF}\, |C_{\rm s}^{(j)}(B)|^{2}\, k_{\rm dis}}
{k_{\rm TF}\, |C_{\rm s}^{(j)}(B)|^{2}+k_{\rm dis}}
= k_{\rm SF}\, Y_{\rm fiss}^{(t)}(B),
\label{eq:kfisstB}
\end{eqnarray}
and
\begin{eqnarray}
Y_{\rm fiss}^{(t)}(B)=\sum_{j=1}^9
\frac{|C_{\rm s}^{(j)}(B)|^{2}}
{1+k_{\rm Td}\, |C_{\rm s}^{(j)}(B)|^{2}},
\qquad
k_{\rm Td}\equiv \frac{k_{\rm TF}}{k_{\rm dis}} .
\label{eq:YfisstB}
\end{eqnarray}
Thus, the magnetic field affects the fast overall singlet fission rate constant including the first dissociation of the triplet pair 
through redistribution of the singlet characters $C_{\rm s}^{(j)}(B)$. 
The dissociation of the triplet pair competes with triplet fusion. 
Equation~(\ref{eq:kfisstB}) makes this explicit: the transient triplet-fission-to-fusion branching rate $k_{\rm fiss}^{(t)}(B)$ enters $k_{\rm tr}(B)$ only through the weighted sum $Y_{\rm fiss}^{(t)}(B)$ over the field-dependent singlet characters $|C_{\rm s}^{(j)}(B)|^2$, since the separation (dissociation) of a neighboring triplet-exciton pair competes with triplet fusion for each spin state $j$. 
This confirms quantitatively that the entire magnetic-field dependence of $k_{\rm tr}(B)$, and hence of the short-time emission, originates from the fusion-versus-dissociation branching captured by $Y_{\rm fiss}^{(t)}(B)$, not from singlet fission.

\subsection{Long-time region: diffusion-controlled geminate recombination tail amplitude}
\label{subsec:criterion_slow}

In three dimensions, the regeneration kernel [Eq.~(\ref{eq:FPTD})] takes the explicit form
\begin{eqnarray}
\hat{R}(s)=\frac{a}{a+b}\exp\!\left(-b\sqrt{s/D}\right).
\label{eq:regK3D}
\end{eqnarray}
For reference, in one dimension it simplifies to
\begin{eqnarray}
\hat{R}(s)=\exp\!\left(-b\sqrt{s/D}\right),
\end{eqnarray}
and satisfies $\hat{R}(s\to 0)=1$ due to recurrence of the random walk \cite{Feller_71}.
In this case, all separated pairs eventually re-encounter and recombine via diffusion-controlled geminate processes, so that no escape channel exists and the final state is uniquely determined by geminate recombination. This one-dimensional limit therefore provides a minimal setting in which the mechanism of trajectory crossing can be isolated without bulk diffusion-controlled recombination.

The small-$s$ expansion of Eq.~(\ref{eq:regK3D}) produces a non-analytic $\sqrt{s}$ term in Eq.~(\ref{eq:LaplaceS}). Introducing the renormalized dissociation rate
\begin{eqnarray}
k_{\rm dis}^\dagger = k_{\rm dis}\frac{b}{a+b},
\label{eq:kdisdagger}
\end{eqnarray}
one obtains, for small $s$ \cite{seki2018},
\begin{eqnarray}
\frac{\hat{S}(s)}{S(0)}
\simeq
\left[
k_{\rm rad}+k_{\rm fiss}(B)
+\sum_j Y_{\rm fus}^{(j)}(B)\, k_{\rm fiss}^{(j)}(B)\, a\sqrt{s/D}
\right]^{-1},
\label{eq:LaplaceSappr}
\end{eqnarray}
where $k_{\rm fiss}(B)=\sum_j k_{\rm fiss}^{(j)}(B)$ and
\begin{eqnarray}
k_{\rm fiss}^{(j)}(B)
&=&
\frac{k_{\rm SF} |C_{\rm s}^{(j)}(B)|^2 k_{\rm dis}^\dagger}
{k_{\rm TF} |C_{\rm s}^{(j)}(B)|^2+k_{\rm dis}^\dagger},
\label{eq:sfj}\\
Y_{\rm fus}^{(j)}(B)
&=&
\frac{k_{\rm TF} |C_{\rm s}^{(j)}(B)|^2}
{k_{\rm TF} |C_{\rm s}^{(j)}(B)|^2+k_{\rm dis}^\dagger}.
\label{eq:Ysj}
\end{eqnarray}

Expanding further in powers of $a\sqrt{s/D}$ and performing the inverse Laplace transformation, the $\sqrt{s}$ term yields the long-time power-law decay. In three dimensions, the asymptotic behavior can be written as \cite{seki2018}
\begin{eqnarray}
S (t)\approx \frac{k_{\rm TF}}{k_{\rm SF} k_{\rm dis}^\dagger}\,
\frac{a}{2\sqrt{\pi D}}\,
\frac{A_{\rm n}(B)}{t^{3/2}},
\label{eq:Stail_An}
\end{eqnarray}
where
\begin{eqnarray}
A_{\rm n}(B)=\sum_j
\frac{ |C_{\rm s}^{(j)}(B)|^{4} }
{\left(1+k_{\rm Td}^\dagger\, |C_{\rm s}^{(j)}(B)|^{2}\right)^2
\left(k_{\rm rS}+Y_{\rm fiss}^\dagger(B)\right)^2},
\label{eq:AnB}
\end{eqnarray}
and
\begin{eqnarray}
Y_{\rm fiss}^\dagger(B)=\sum_j
\frac{ |C_{\rm s}^{(j)}(B)|^{2} }
{1+k_{\rm Td}^\dagger\, |C_{\rm s}^{(j)}(B)|^{2}} .
\label{eq:YfissdagB}
\end{eqnarray}
Here $k_{\rm Td}^\dagger=k_{\rm TF}/k_{\rm dis}^\dagger$ and $k_{\rm rS}=k_{\rm rad}/k_{\rm SF}$. 
This diffusion-controlled geminate recombination tail acts as an effective \emph{slow} contribution associated with repeated re-encounters of the geminate pair, and the asymptotic amplitude depends on the magnetic field through $C_{\rm s}^{(j)}(B)$.

In dimensions $d\le 2$, dissociated geminate triplet pairs inevitably re-encounter at long times due to recurrence \cite{Feller_71}.
Consequently, recombination proceeds exclusively through diffusion-controlled geminate processes of the originally correlated pair, and no population remains indefinitely separated. In this limit, $\hat{R}(s\to 0)=1$. Defining $\hat{F}_{\rm pt}(s)=1-\hat{R}(s)$ and considering Eq.~(\ref{eq:LaplaceS}) for $\hat{F}_{\rm pt}(s)\to 0$ as $s\to 0$, Eq.~(\ref{eq:FPTD}) implies $s+k_{\rm dis}\hat{F}_{\rm pt}(s)\approx k_{\rm dis}\hat{F}_{\rm pt}(s)$ in the asymptotic limit. Equation~(\ref{eq:LaplaceS}) then reduces to
\begin{align}
\frac{\hat{S}(s)}{S(0)}
&\approx 
\left[
k_{\rm rad}+k_{\rm SF}
-\sum_{j=1}^9
\frac{k_{\rm SF}k_{\rm TF} |C_{\rm s}^{(j)}(B)|^{4}}
{k_{\rm TF} |C_{\rm s}^{(j)}(B)|^2+k_{\rm dis}\hat{F}_{\rm pt}(s)}
\right]^{-1}\\
&=
\left[
k_{\rm rad}+k_{\rm SF}
-\sum_{j=1}^9
\frac{k_{\rm SF}|C_{\rm s}^{(j)}(B)|^{2}}
{1+k_{\rm dis}\hat{F}_{\rm pt}(s)/\left(k_{\rm TF} |C_{\rm s}^{(j)}(B)|^2\right)}
\right]^{-1}\\
&\approx 
\left[k_{\rm rad}\left(
1+
9 \frac{k_{\rm SF} k_{\rm dis}}{k_{\rm rad} k_{\rm TF}} 
\hat{F}_{\rm pt}(s)
\right) \right]^{-1},
\label{eq:LaplaceS12}
\end{align}
which shows explicitly that the leading asymptotic term becomes independent of $C_{\rm s}^{(j)}(B)$ and hence of the magnetic field within the Johnson--Merrifield framework \cite{seki2018}.

In three dimensions, by contrast, the random walk is transient and a finite fraction of dissociated pairs escapes geminate recombination, leading to a residual population of separated triplets. 
These triplets subsequently relax through diffusion-controlled bulk recombination processes, in which recombination occurs independently of the original pairing.

\section{Formulation of generalized Mpemba trajectory-crossing criterion}
\label{subsec:trajectorycriterion}

We identify the generalized Mpemba effect with trajectory crossing toward a common final state. A sufficient condition for such crossing is that the ordering of the trajectories reverses between the early- and late-time regimes. Specifically, crossing occurs when
\begin{eqnarray}
k_{\rm tr}(B_2) > k_{\rm tr}(B_1),
\label{eq:criterion_fast}
\end{eqnarray}
where $k_{\rm tr}(B)$ is given by Eq.~(\ref{eq:ktrB}), and
\begin{eqnarray}
A_{\rm n}(B_2) > A_{\rm n}(B_1),
\label{eq:criterion_slow}
\end{eqnarray}
where $A_{\rm n}(B)$ is given by Eq.~(\ref{eq:AnB}). The opposite ordering,
\begin{eqnarray}
k_{\rm tr}(B_2) < k_{\rm tr}(B_1), \qquad A_{\rm n}(B_2) < A_{\rm n}(B_1),
\label{eq:criterion2}
\end{eqnarray}
also leads to trajectory crossing.

Denoting the excited singlet populations at magnetic fields $B_2$ and $B_1$ by $S_{B_2}(t)$ and $S_{B_1}(t)$, respectively, these conditions imply that
\begin{eqnarray}
\Delta S(t)\equiv S_{B_2}(t)-S_{B_1}(t)
\end{eqnarray}
changes sign. Hence, there exists a crossing time $t_\times$ such that
\begin{eqnarray}
S_{B_2}(t_\times)=S_{B_1}(t_\times).
\end{eqnarray}

Thus, trajectory crossing arises from competition between an effective fast decay and a diffusion-controlled geminate-recombination power-law contribution, whose relative weights are redistributed by the magnetic field while the final state remains unchanged. This mechanism directly corresponds to the short- and long-time contributions identified in the preceding section.

This criterion provides a minimal and physically transparent condition for the emergence of the generalized Mpemba effect in diffusion-controlled geminate-recombination systems. It is not intended to be necessary in all parameter regimes; rather, it gives a sufficient condition when the relaxation can be separated into an effective transient decay and an asymptotic diffusion-controlled tail.

\subsection{Zero- and weak-field region}
\label{sec:criterion_field}

In the zero-field limit (see \ref{app:singlet_character}), three of the nine triplet-pair spin states carry singlet character when the Hamiltonian is invariant under exchange of the two triplets \cite{merrifield1968,johnson1970,merrifield1971}.
the magnetic field strength increases, the singlet character is redistributed among the remaining states through mixing of the zero-field eigenstates.

In a simplified treatment, this redistribution can be approximated by assuming equal singlet character among the participating triplet-pair spin states. Specifically, we take $|C_{\rm s}^{(j)}(B)|^2=1/N_{\rm s}$ for $N_{\rm s}$ states, where $N_{\rm s}=3$ in the zero-field limit and $N_{\rm s}=9$ in the weak-field region \cite{merrifield1968,johnson1970,merrifield1971,steiner1989}.

In the regime where the relaxation trajectories obtained at zero and weak magnetic fields cross, the crossing condition [Eq. (\ref{eq:criterion_slow})] can be expressed as \cite{seki2018}
\begin{eqnarray}
\frac{k_{\rm TF}}{k_{\rm dis}}\, 
\frac{k_{\rm rad}/k_{\rm SF}}{1+k_{\rm rad}/k_{\rm SF}}
\geq
9 \frac{b}{a+b}.
\label{eq:Mpemba_threshold_low}
\end{eqnarray}
This condition shows explicitly that the triplet-fusion rate constant must be sufficiently larger than the dissociation rate constant for trajectory crossing to occur.

\subsection{High-field region}
\label{sec:criterion_highfield}

Here, we consider the high-field-limit representation (see \ref{app:high_field_singlet_character}) \cite{johnson1970,merrifield1971,steiner1989}
\begin{eqnarray}
|C_{\rm s}^{(1)}(B)|^2 = \frac{1}{3}, \qquad |C_{\rm s}^{(2)}(B)|^2 = \frac{2}{3},
\label{eq:Cs_highfield}
\end{eqnarray}
with the remaining states having negligible singlet character. In this limit, the field-induced redistribution of $C_{\rm s}^{(j)}(B)$ modifies both the transient decay and the amplitude of the diffusion-controlled geminate-recombination tail.

In the regime where the relaxation trajectories obtained at zero and high magnetic fields cross, the crossing condition [Eq. (\ref{eq:criterion2})] can be expressed as \cite{seki2018}
\begin{eqnarray}
\frac{k_{\rm TF}}{k_{\rm dis}} > \frac{3\sqrt{2}}{2} \frac{b}{a+b},
\label{eq:Mpemba_threshold}
\end{eqnarray}
which again shows that the triplet-fusion rate constant must be sufficiently larger than the dissociation rate constant for trajectory crossing to occur.

\section{Results}

\begin{figure}
\centerline{
\includegraphics[width=1\columnwidth]{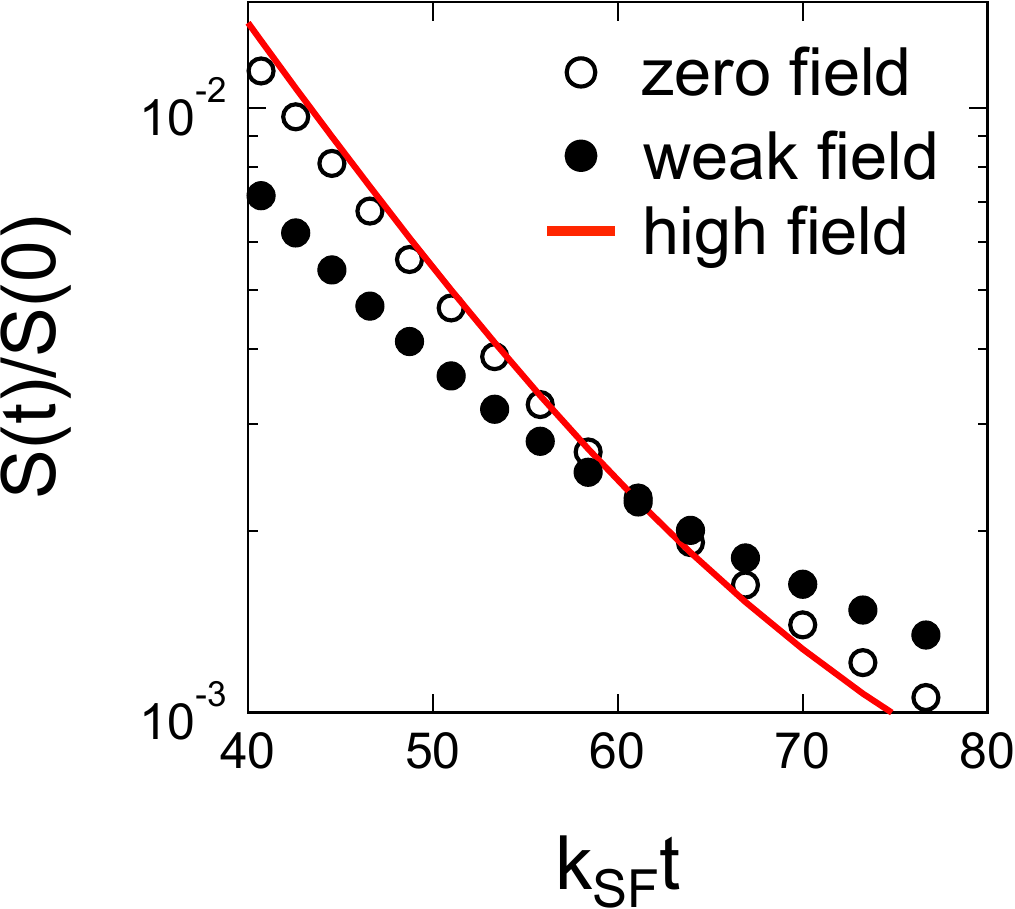}
}
\caption{Fluorescence relaxation trajectories exhibiting the generalized Mpemba effect. 
$S(t)/S(0)$ is proportional to normalized delayed fluorescence intensity.
The black open circles, black filled circles, and red thick solid line represent the relaxation trajectories in the absence of a magnetic field ($N_{\rm s}=3$), 
in the weak-field region ($N_{\rm s}=9$), and in the high-field limit [Eq.~(\ref{eq:Cs_highfield})], respectively. 
The trajectories cross one another. 
The parameter values are 
$k_{\rm Td}=80$, $k_{\rm rS}=0.1$, $k_{\rm SF}/k_{\rm dis}=1$, $D/(a^2k_{\rm SF})=1$, and 
$a=b$.
} 
\label{fig:Mpemba_cross}
\end{figure}

We begin by examining the magnetic-field dependence of fluorescence decay kinetics in triplet-fusion systems, focusing on the diffusion-controlled geminate-recombination regime. Figure~\ref{fig:Mpemba_cross} presents representative results obtained by numerical inverse Laplace transformation of Eq.~(\ref{eq:LaplaceS}) using the Stehfest method \cite{Stehfest1970_47,Stehfest1970_624}. The parameter values are slightly modified from those used in Fig.~4 of Ref.~\cite{seki2018}, and the time window is restricted to $40 \le t \le 80$ to resolve the trajectory crossing more clearly.

The black open and filled circles, together with the solid red line, represent relaxation of the same observable: the singlet population, which is proportional to the delayed fluorescence intensity. The trajectories follow pulsed excitation of the singlet excited state and relax toward a common final state, differing only in the external control parameter $B$.

In the absence of a magnetic field, singlet fission generates triplet pairs that undergo both dissociation and diffusion-controlled geminate recombination. 
While dissociation leads to a loss of spin correlation, separated triplet pairs can repeatedly re-encounter via diffusion, resulting in regeneration of the triplet-pair state and subsequent triplet fusion. 
This competition between dissociation and diffusion-controlled geminate recombination gives rise to delayed fluorescence, non-exponential relaxation dynamics, and ultimately the observed trajectory crossing.

With increasing magnetic field, spin mixing among triplet sublevels modifies the kinetics of triplet fusion, resulting in either enhanced or suppressed regeneration of the singlet state at intermediate and long times. 

As shown in Fig.~\ref{fig:Mpemba_cross}, the black filled circles (weak-field case) initially decay faster than the black open circles (zero-field case), subsequently cross them, and eventually relax more slowly toward the common final state. 

Although the lowest triplet sublevel decreases with increasing magnetic field via Zeeman splitting, which would naively facilitate relaxation, the delayed fluorescence decays more slowly at finite but weak magnetic fields. 
This counterintuitive behavior cannot be understood within a purely energetic picture, but instead reflects the kinetic competition between dissociation and diffusion-controlled geminate recombination. 
We therefore interpret the magnetic-field-dependent crossing in Fig.~\ref{fig:Mpemba_cross} as a realization of the generalized Mpemba effect in a spin-dependent photokinetic system.

In contrast, the red line (high-field case) initially decays more slowly than the zero-field trajectory, subsequently crosses it, and eventually relaxes faster toward the common final state. 

Consequently, the relaxation trajectories associated with different magnetic fields exhibit a nontrivial reordering in time, leading to the trajectory crossing observed in Fig.~\ref{fig:Mpemba_cross}.
 We also note that the most significant Mpemba-like crossing is obtained between the weak- and high-field trajectories in Fig.~\ref{fig:Mpemba_cross}.

\subsection{Mode competition viewpoint}

A more transparent understanding of trajectory crossing can be obtained by analyzing the relaxation dynamics in terms of competing kinetic contributions separated by pair dissociation.

At short times, before pair dissociation, the fluorescence decay reflects local spin-selective processes, including rapid singlet fission and triplet-pair fusion, leading to a fast decrease in the singlet population. After pair dissociation, the relaxation is governed by diffusion-controlled geminate recombination of the separated triplet pair. This process gives rise to a power-law decay of the fluorescence intensity, reflecting the probability of geminate re-encounter between diffusing triplet excitons.

From this viewpoint, the short- and long-time behaviors can be regarded as distinct kinetic contributions: an effective fast contribution associated with local singlet fission and triplet-pair fusion before pair dissociation, and a slow, non-exponential contribution associated with diffusion-controlled geminate recombination after pair dissociation. The magnetic field modifies the relative weights of these contributions without altering the final state, thereby redistributing the relaxation pathways over time.

This interpretation highlights that the generalized Mpemba effect in this system does not originate from a single relaxation rate constant, but from competition and redistribution among multiple kinetic contributions across different time scales. 
In particular, 
trajectory crossing arises when the magnetic field redistributes the transient decay rate constant and the long-time power-law amplitude so that the ordering of trajectories reverses between early and late times.

The slow power-law decay originates from diffusion-controlled geminate recombination. 
Within the Johnson--Merrifield framework, spin mixing in the long-time regime is affected by the magnetic field only when the escape probability is finite. 
Therefore, in the recurrent diffusion regime (dimensions $\le 2$), trajectory crossing does not generally arise within this framework, where off-diagonal elements of the spin Hamiltonian are neglected. 
In contrast, crossing becomes possible in transient diffusion (dimensions $d>2$), where the escape probability is finite, provided that the triplet fusion rate constant is sufficiently larger than the dissociation rate constant [Eqs.~(\ref{eq:Mpemba_threshold_low}) and (\ref{eq:Mpemba_threshold})].

\subsection{Phase diagram}
\label{subsec:phase_diagram}

\begin{figure}
\centerline{
\includegraphics[width=0.7\columnwidth]{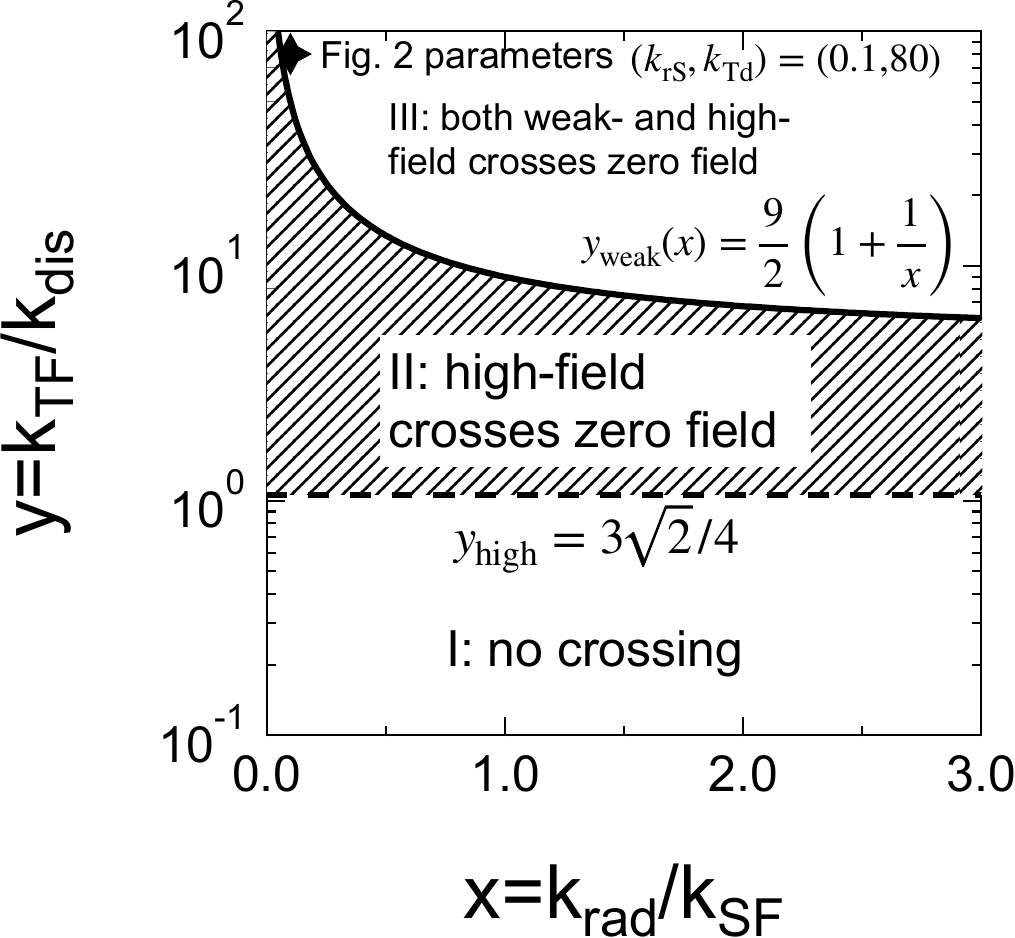}
}
\caption{ Phase diagram in the $(x,y)=(k_{\rm rad}/k_{\rm SF},\,k_{\rm TF}/k_{\rm dis})$ plane for symmetric encounter geometry ($a=b$). Region I: neither the weak- nor the high-field trajectory crosses the zero-field trajectory. Region II: only the high-field trajectory crosses the zero-field trajectory [Eq.~(\ref{eq:xhigh_boundary})]. Region III: both the weak- and high-field trajectories cross the zero-field trajectory [Eq.~(\ref{eq:xweak_boundary})]. The parameters $(k_{\rm rS},k_{\rm Td})=(0.1,80)$ used in Fig.~\ref{fig:Mpemba_cross}, marked in the panel, lie within region III.}
\label{fig:phase_diagram}
\end{figure}

Equations~(\ref{eq:Mpemba_threshold_low}) and~(\ref{eq:Mpemba_threshold}) depend on the encounter geometry only through $b/(a+b)$. Setting $a=b$, i.e., taking the initial separation $a+b$ to be twice the encounter distance $a$, gives $b/(a+b)=1/2$, and the two conditions reduce to
\begin{eqnarray}
\frac{k_{\rm TF}}{k_{\rm dis}}\,\frac{k_{\rm rad}/k_{\rm SF}}{1+k_{\rm rad}/k_{\rm SF}} \geq \frac{9}{2}
\label{eq:Mpemba_threshold_low_ab}
\end{eqnarray}
for the zero-versus-weak-field crossing, and
\begin{eqnarray}
\frac{k_{\rm TF}}{k_{\rm dis}} > \frac{3\sqrt{2}}{4} \approx 1.06
\label{eq:Mpemba_threshold_ab}
\end{eqnarray}
for the zero-versus-high-field crossing. Denoting $y\equiv k_{\rm TF}/k_{\rm dis}$ and $x\equiv k_{\rm rad}/k_{\rm SF}$, Eq.~(\ref{eq:Mpemba_threshold_low_ab}) defines the boundary curve
\begin{eqnarray}
y_{\rm weak}(x) = \frac{9}{2}\left(1+\frac{1}{x}\right),
\label{eq:xweak_boundary}
\end{eqnarray}
above which the zero- and weak-field trajectories cross, while Eq.~(\ref{eq:Mpemba_threshold_ab}) defines the $x$-independent boundary
\begin{eqnarray}
y_{\rm high} = \frac{3\sqrt{2}}{4},
\label{eq:xhigh_boundary}
\end{eqnarray}
above which the zero- and high-field trajectories cross. Because $y_{\rm weak}(x) > 9/2 > y_{\rm high}$ for every $x>0$, the $(x,y)$ plane divides into three regions [Fig.~\ref{fig:phase_diagram}]: (I) $y<y_{\rm high}$, where neither pair of trajectories crosses; (II) $y_{\rm high}<y<y_{\rm weak}(x)$, where only the zero- and high-field trajectories cross; and (III) $y>y_{\rm weak}(x)$, where both pairs of trajectories cross. Because typical singlet-fission systems have $k_{\rm SF}\gg k_{\rm rad}$, so that $x\ll1$, the weak-field boundary diverges as $x\to0$ [Eq.~(\ref{eq:xweak_boundary})], making region III comparatively difficult to reach in practice, whereas the high-field boundary $y_{\rm high}$ is fixed and modest; region II, where only the high-field crossing is realized, therefore occupies most of the physically relevant parameter space. 
The crossing of relaxation trajectories observed experimentally indicates that $y$ might be higher than $y_{\rm high}$ \cite{burdett2013,piland2013}. 
The parameters used in Fig.~\ref{fig:Mpemba_cross} below, $(k_{\rm rS},k_{\rm Td})=(0.1,80)$, satisfy $y_{\rm weak}(0.1)=49.5<80$ and are therefore located in region III, consistent with the simultaneous weak- and high-field crossings shown there [marked point, Fig.~\ref{fig:phase_diagram}].

\subsection{A possible influence of protocols}
Several protocols, which explicitly take into account the time needed to prepare the initial non-equilibrium state, have recently been proposed to induce or classify generalized Mpemba effects, including the Pontus protocol (a two-step protocol) \cite{pontus2025} and the Descartes protocol (a single-step protocol using three reservoirs). \cite{descartes2026} 
Here, we study briefly whether an analogous protocol-based description applies to the present spin-selective kinetic system.  

Suppose that the field is applied only up to a switching time $t_{\rm sw}$ comparable to the onset of the triplet-fusion-dominated regime (of order tens of nanoseconds), and switched off for $t>t_{\rm sw}$. 
However, this switching protocol would not emphasize the crossing; if anything, it works against it. 
The amplitude $A_{\rm n}(B)$ [Eq.~(\ref{eq:AnB})] is not a fixed weight but the asymptotic outcome of the geminate pair repeatedly dissociating and re-encountering under a persistently applied field $B$: 
it presumes that every re-encounter event is governed by the same $C_{\rm s}^{(j)}(B)$ with the same $B$. 
However, we roughly estimate the effect of switching off the field ($B$) at $t_{\rm sw}$, 
and assume that every re-encounter event for $t>t_{\rm sw}$ is instead governed by $C_{\rm s}^{(j)}(0)$. 
As a result, for $t\gg t_{\rm sw}$ the tail can be approximately characterized by the same asymptotic amplitude as $A_{\rm n}(0)$, 
common to both the switched and the always-zero-field trajectories; 
the two trajectories can then differ only by an overall effect set by the history accumulated up to $t_{\rm sw}$. 
Switching the field off instead converts the problem into a comparison of 
two different initial conditions (the field-on and always-zero histories at $t_{\rm sw}$) relaxing under identical zero-field dynamics thereafter -- a legitimate but generally weaker realization of the effect, 
governed by the overlap of these histories with the zero-field slow mode rather than by Eqs.~(\ref{eq:criterion_fast})--(\ref{eq:criterion2}).

\subsection{Distance measure}

We note that  the distance measure from the final state in the generalized Mpemba effect \cite{distance_measure_granular2023,thermomajorization2025} 
is formulated directly in terms of the observable $S(t)$ (equivalently, the delayed-fluorescence intensity), 
which disappears when all excitons convert to singlet ground states. 
The crossing criterion in Eqs.~(\ref{eq:criterion_fast})--(\ref{eq:criterion2}) is obtained for $S(t)$. 
Because the final state here is independent of the magnetic field, any monotonic distance measure constructed from $S(t)$ itself would exhibit the same crossing behavior as $S(t)$; the crossing property is therefore robust with respect to the choice among such observable-based measures. 

Here, we study whether the same crossing persists under distance measures defined on 
the total associated triplet-pair population $C(t)=\sum_j C^{(j)}(t)$, rather than on the singlet population alone.  
The crossing discussed above is a property of the singlet population $S(t)$, the quantity proportional to the observed delayed fluorescence intensity. 
Laplace transform of Eq.~(\ref{eq:Cd}) gives, for each spin state $j$,
\begin{eqnarray}
\hat C^{(j)}(s) = \frac{k_{\rm SF}|C_{\rm s}^{(j)}(B)|^2}{s+k_{\rm TF}|C_{\rm s}^{(j)}(B)|^2+k_{\rm dis}\left(1-\hat R(s)\right)}\,\hat S(s),
\label{eq:Chat_j}
\end{eqnarray}
so that the total associated triplet-pair population is $\hat C(s)=\sum_{j=1}^9 \hat C^{(j)}(s)$. Numerically inverting $\hat C(s)$ by the Stehfest algorithm, using the parameters of Fig.~\ref{fig:Mpemba_cross}, shows that the ordering of the zero-, weak-, and high-field trajectories of $C(t)$ remains $C_{\rm high}(t)<C_{\rm zero}(t)<C_{\rm weak}(t)$ throughout the entire range examined ($10^{-2}\lesssim k_{\rm SF}t\lesssim 3\times10^3$), with no crossing, in contrast to the crossing exhibited by $S(t)$ near $k_{\rm SF}t\approx 66$ 
(See Fig. \ref{fig:Mpemba_cross}). 

\begin{figure}
\centerline{
\includegraphics[width=1\columnwidth]{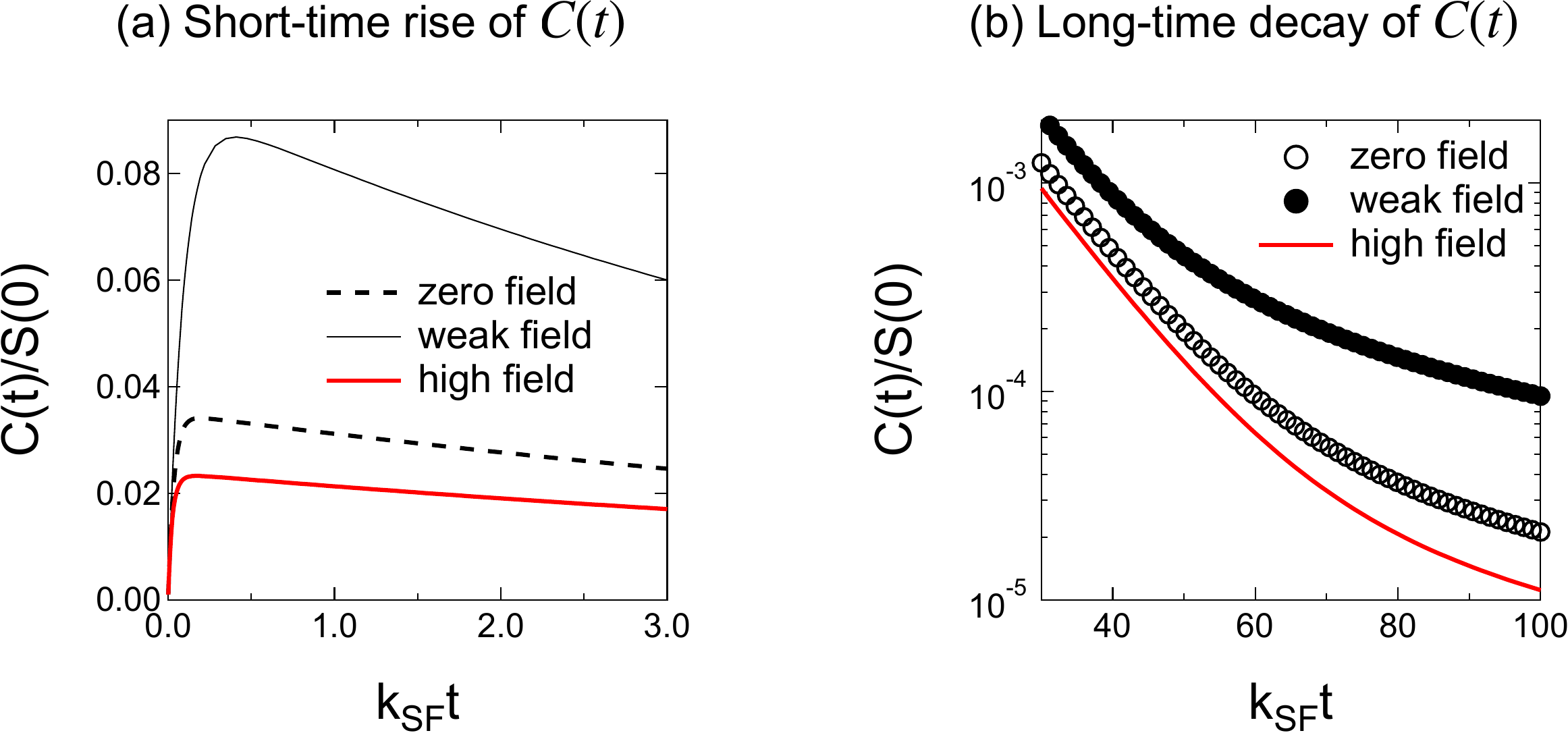}
}
\caption{ Rise and decay of the total associated triplet-pair population $C(t)=\sum_j C^{(j)}(t)$, obtained by numerical inverse Laplace transformation using the same parameters as Fig.~\ref{fig:Mpemba_cross} ($k_{\rm Td}=80$, $k_{\rm rS}=0.1$, $k_{\rm SF}/k_{\rm dis}=1$, $D/(a^2k_{\rm SF})=1$, $a=b$). (a) Short-time regime: $C(t)$ rises from $C(0)=0$ to a maximum whose height and timing follow the short-time depletion rate of $S(t)$ (weak field: highest, latest peak; high field: lowest, earliest peak). (b) Long-time regime, including the time window of Fig.~\ref{fig:Mpemba_cross}: the ordering $C_{\rm high}<C_{\rm zero}<C_{\rm weak}$ set at short times persists with no crossing.}
\label{fig:SvsC}
\end{figure}

Since $C(0)=0$ for every field, $C(t)$ must rise from zero, reach a maximum, and subsequently decay [Fig.~\ref{fig:SvsC}(a)]. Whichever field depletes $S(t)$ fastest at short times diverts the most flux into $C(t)$, producing the highest and latest-occurring peak: the weak-field case, which depletes $S(t)$ fastest [Fig.~\ref{fig:Mpemba_cross}], reaches the highest peak of $C(t)$, while the high-field case, which depletes $S(t)$ most slowly, reaches the lowest peak. This peak ordering, $C_{\rm high}<C_{\rm zero}<C_{\rm weak}$, is set within $k_{\rm SF}t\lesssim1$ 
and is identical to the common ordering between $S(t)$ and $C(t)$ that persists to long times [Figs.~\ref{fig:Mpemba_cross} and \ref{fig:SvsC}(b)]. This is the origin of the absence of crossing in $C(t)$: unlike $S(t)$, whose short-time ordering is the reverse of its long-time ordering, $C(t)$'s ordering is fixed once and for all by the initial flux redistribution and never needs to reverse.

This shows that the generalized Mpemba effect discussed in this work is therefore a property specific to the fluorescence observable $S(t)$ 
in this non-Markovian relaxation process, 
rather than a generic feature of every dynamical variable in the model.

As a further check on the initial-condition treatment adopted throughout this work, \ref{app:reverse_ic} verifies the underlying time-scale-separation assumption directly, by comparing the crossing obtained from $S(0)$ against that obtained by imposing each field's post-mixing triplet-pair distribution as the initial condition.

\section{Conclusion}

In this work, we have revisited magnetic-field-dependent delayed fluorescence in triplet-fusion systems from the perspective of anomalous nonequilibrium relaxation and demonstrated that the observed crossing of relaxation trajectories constitutes a realization of the generalized Mpemba effect in a spin-dependent photokinetic system.

The fluorescence relaxation trajectories at different magnetic fields correspond to the relaxation of the same observable toward a common final state and differ only in the external control parameter. The trajectory crossing observed in such systems therefore directly satisfies the defining feature of the generalized Mpemba effect in its modern formulation. This behavior cannot be explained by the simple energetic expectation that lowering the triplet sublevel should enhance relaxation; instead, it arises from the underlying kinetic structure of the system.

Using the Johnson--Merrifield framework extended to include diffusion-controlled geminate recombination, we identified two qualitatively distinct contributions governing the relaxation dynamics: an effective fast contribution associated with local singlet fission and triplet-pair fusion before pair dissociation, and a slow, non-exponential contribution associated with diffusion-controlled geminate recombination after pair dissociation. The magnetic field modifies the singlet character of triplet-pair states, thereby altering the relative weights of the fast and slow contributions without affecting the final state, in which all triplets are eventually converted back to the singlet ground state.

Within this framework, we formulated a generalized Mpemba criterion based on mode competition. Trajectory crossing arises when the magnetic field changes the effective transient decay and the amplitude of the long-time diffusion-controlled geminate-recombination contribution in a way that reverses the ordering of relaxation trajectories.

An important aspect of the present system is that the slow contribution originates from diffusion-controlled geminate recombination, resulting in a power-law decay rather than a discrete exponential mode. The generalized Mpemba effect discussed here therefore extends the conventional finite-mode picture and demonstrates that anomalous relaxation can emerge from competition between qualitatively different kinetic processes.

For theoretical clarity, it is instructive to consider lower-dimensional cases ($d \leq 2$), in which the recurrence property of diffusion eliminates the escape channel and the final state is uniquely determined by geminate recombination. In this regime, the long-time power-law decay becomes independent of spin mixing within the Johnson--Merrifield framework and is therefore insensitive to the magnetic field. In contrast, in three dimensions, a finite fraction of triplet pairs escapes geminate recombination and subsequently relaxes through diffusion-controlled bulk recombination.

The present results establish a unified physical picture in which geminate diffusion, spin dynamics, and kinetic competition jointly give rise to anomalous relaxation. More broadly, they suggest that generalized Mpemba effects may be a generic feature of systems with separated time scales and non-exponential relaxation, particularly in non-Markovian diffusion-controlled recombination processes and strongly interacting intermediate-state kinetics.


\appendix
\setcounter{figure}{0}

\section{Singlet character in the Cartesian triplet basis}
\label{app:singlet_character}

Here we summarize the origin of the equal-singlet-character approximation used in Sec.~\ref{sec:criterion_field}. Consider two triplet excitons, denoted by $A$ and $B$, each with spin quantum number $S=1$. We write the single-triplet spin states in the spherical basis as
\begin{eqnarray}
|+\rangle \equiv |1,+1\rangle,\qquad
|0\rangle \equiv |1,0\rangle,\qquad
|-\rangle \equiv |1,-1\rangle .
\end{eqnarray}
The nine triplet-pair product states span the direct-product space of two spin-one objects. Coupling the two triplets gives total spin states with $J=0,1,2$. The total spin-singlet state of the two triplets is
\begin{eqnarray}
|J=0,M=0\rangle
=
\frac{1}{\sqrt{3}}
\left(
|0\rangle_A |0\rangle_B
-
|+\rangle_A |-\rangle_B
-
|-\rangle_A |+\rangle_B
\right),
\label{eq:appendix_singlet_pair_pm}
\end{eqnarray}
apart from an arbitrary overall phase convention.

The singlet character of a triplet-pair spin state $|\psi_j\rangle$ is defined as the squared projection onto the total spin-singlet state,
\begin{eqnarray}
|C_{\rm s}^{(j)}|^2
\equiv
|\langle J=0,M=0|\psi_j\rangle|^2 .
\label{eq:appendix_singlet_character_definition}
\end{eqnarray}

The same total spin-singlet state can be expressed in the Cartesian triplet basis, $\{|T_x\rangle,|T_y\rangle,|T_z\rangle\}$. With the convention
\begin{eqnarray}
|T_x\rangle &=& \frac{1}{\sqrt{2}}\left(|-\rangle-|+\rangle\right),
\\
|T_y\rangle &=& \frac{i}{\sqrt{2}}\left(|+\rangle+|-\rangle\right),
\\
|T_z\rangle &=& |0\rangle ,
\end{eqnarray}
one obtains
\begin{eqnarray}
|T_x\rangle_A |T_x\rangle_B
+
|T_y\rangle_A |T_y\rangle_B
+
|T_z\rangle_A |T_z\rangle_B
=
|0\rangle_A |0\rangle_B
-
|+\rangle_A |-\rangle_B
-
|-\rangle_A |+\rangle_B .
\label{eq:appendix_cartesian_identity}
\end{eqnarray}
Thus, Eq.~(\ref{eq:appendix_singlet_pair_pm}) can equivalently be written as
\begin{eqnarray}
|J=0,M=0\rangle
=
\frac{1}{\sqrt{3}}
\left(
|T_x\rangle_A |T_x\rangle_B
+
|T_y\rangle_A |T_y\rangle_B
+
|T_z\rangle_A |T_z\rangle_B
\right).
\label{eq:appendix_singlet_pair_xyz}
\end{eqnarray}

Equation~(\ref{eq:appendix_singlet_pair_xyz}) shows explicitly that, in the zero-field Cartesian representation, the total spin-singlet state is distributed equally among the three product states $|T_x\rangle_A |T_x\rangle_B$, $|T_y\rangle_A |T_y\rangle_B$, and $|T_z\rangle_A |T_z\rangle_B$. If these three states are taken as the participating triplet-pair spin states, the singlet character of each state is therefore
\begin{eqnarray}
|C_{\rm s}^{(j)}|^2=\frac{1}{3}
\qquad (j=1,2,3),
\label{eq:appendix_zero_field_character}
\end{eqnarray}
while the remaining six triplet-pair states have no singlet character in this simplified zero-field limit.

When a weak magnetic field mixes the zero-field eigenstates, the singlet character can be redistributed among the triplet-pair spin states. In the simplified treatment used in the main text, this redistribution is approximated as uniform over $N_{\rm s}$ participating states:
\begin{eqnarray}
|C_{\rm s}^{(j)}(B)|^2=\frac{1}{N_{\rm s}},
\qquad
\sum_{j=1}^{N_{\rm s}} |C_{\rm s}^{(j)}(B)|^2=1 .
\label{eq:appendix_uniform_character}
\end{eqnarray}
This gives $N_{\rm s}=3$ in the zero-field limit and $N_{\rm s}=9$ in the weak-field region. The approximation preserves the normalization of the total singlet character while replacing the detailed field-dependent spin mixing by an equal redistribution over the participating triplet-pair spin states.

\section{High-field-limit singlet character}
\label{app:high_field_singlet_character}

Here we summarize the origin of the high-field-limit representation used in Sec.~\ref{sec:criterion_highfield}. The high-field-limit projection is used here as an idealized spin-statistical representation within the Johnson--Merrifield framework. It gives the singlet-character weights of selected high-field basis states through their squared overlaps with the total spin-singlet state. This projection does not describe the full time-dependent spin dynamics of correlated triplet pairs, for which coherences, exchange interaction, dipolar coupling, molecular orientation, and spin relaxation may be important \cite{yago_16}.

In a strong magnetic field applied along the quantization axis, it is convenient to classify the triplet-pair product states by the total magnetic quantum number
\begin{eqnarray}
M=m_A+m_B .
\end{eqnarray}
Only the three states with $M=0$ are candidates for carrying nonzero singlet character:
\begin{eqnarray}
|0\rangle_A |0\rangle_B,\qquad
|+\rangle_A |-\rangle_B,\qquad
|-\rangle_A |+\rangle_B .
\end{eqnarray}
The states with $M=\pm1$ and $M=\pm2$ have no overlap with the total spin-singlet state.

Within the $M=0$ subspace, we introduce the orthonormal combinations
\begin{eqnarray}
|\Phi_0\rangle
&=&
|0\rangle_A |0\rangle_B ,
\\
|\Phi_+\rangle
&=&
\frac{1}{\sqrt{2}}
\left(
|+\rangle_A |-\rangle_B
+
|-\rangle_A |+\rangle_B
\right),
\\
|\Phi_-\rangle
&=&
\frac{1}{\sqrt{2}}
\left(
|+\rangle_A |-\rangle_B
-
|-\rangle_A |+\rangle_B
\right).
\end{eqnarray}

The state $|\Phi_-\rangle$ is antisymmetric under interchange of the two triplets, $A\leftrightarrow B$. For two spin-one objects, a total-spin state $|J,M\rangle$ has exchange parity $(-1)^J$. Thus, the $J=0$ and $J=2$ subspaces are symmetric, whereas the $J=1$ subspace is antisymmetric. Since $|\Phi_-\rangle$ also has $M=0$, it belongs to the $J=1$, $M=0$ subspace, up to an arbitrary overall phase convention.

This assignment is also consistent with the action of the total spin operator. Defining
\begin{eqnarray}
J_- = J_-^{(A)}+J_-^{(B)},\qquad
J_+ = J_+^{(A)}+J_+^{(B)},
\end{eqnarray}
and using $J_z|\Phi_-\rangle=0$, the identity
\begin{eqnarray}
J^2=J_+J_-+J_z(J_z-1)
\end{eqnarray}
gives $J^2|\Phi_-\rangle=2|\Phi_-\rangle$. The eigenvalue is therefore $J(J+1)=2$, confirming that $J=1$.

Thus, $|\Phi_-\rangle$ is orthogonal to the total spin-singlet state and has no singlet character in the sense of Eq.~(\ref{eq:appendix_singlet_character_definition}). Consequently, only two of the three $M=0$ candidate states, $|\Phi_0\rangle$ and $|\Phi_+\rangle$, have nonzero singlet character.

Using these states, Eq.~(\ref{eq:appendix_singlet_pair_pm}) becomes
\begin{eqnarray}
|J=0,M=0\rangle
=
\frac{1}{\sqrt{3}}|\Phi_0\rangle
-
\sqrt{\frac{2}{3}}|\Phi_+\rangle .
\label{eq:appendix_highfield_decomposition}
\end{eqnarray}
Equation~(\ref{eq:appendix_highfield_decomposition}) then gives
\begin{eqnarray}
|\langle J=0,M=0|\Phi_0\rangle|^2
=
\frac{1}{3},
\qquad
|\langle J=0,M=0|\Phi_+\rangle|^2
=
\frac{2}{3}.
\label{eq:appendix_highfield_overlaps}
\end{eqnarray}

Therefore, in the high-field-limit representation, the singlet character is concentrated in two triplet-pair spin states. Identifying these two participating states as $j=1$ and $j=2$, respectively, one obtains
\begin{eqnarray}
|C_{\rm s}^{(1)}(B)|^2 = \frac{1}{3},
\qquad
|C_{\rm s}^{(2)}(B)|^2 = \frac{2}{3},
\label{eq:appendix_Cs_highfield}
\end{eqnarray}
with the remaining triplet-pair spin states having negligible singlet character in this high-field-limit representation.


\section{Trajectory crossing from field-matched triplet-pair initial conditions}
\label{app:reverse_ic}

The comparisons in the main text all start from the common, field-independent initial condition $S(0)$ with $C^{(j)}(0)=0$, on the grounds that singlet fission followed by fast spin mixing prepares an effectively field-dependent triplet-pair state on a time scale well separated from that of triplet fusion and diffusion-controlled recombination (Sec.~1). Here we check this time-scale-separation assumption directly, by imposing the resulting post-mixing distribution as the initial condition for the triplet-fusion-and-recombination dynamics itself and testing whether the same crossing is recovered as when the full dynamics is evolved from $S(0)$. Concretely, we consider $S(0)=0$ together with a nonzero initial associated triplet-pair population $C^{(j)}(0)=|C_{\rm s}^{(j)}(B)|^2$, i.e., each field's own singlet-character distribution is taken as the initial condition for that field (normalized to unit total population, since $\sum_j|C_{\rm s}^{(j)}(B)|^2=1$).

Repeating the Laplace-transform derivation of Eq.~(\ref{eq:Chat_j}) with a general initial condition $C^{(j)}(0)$, rather than $C^{(j)}(0)=0$, and a general $S(0)$ gives
\begin{align}
\hat C^{(j)}(s) &= \frac{C^{(j)}(0)+k_{\rm SF}|C_{\rm s}^{(j)}(B)|^2\hat S(s)}{D_j(s)},
\nonumber \\
D_j(s)&\equiv s+k_{\rm TF}|C_{\rm s}^{(j)}(B)|^2+k_{\rm dis}\left(1-\hat R(s)\right),
\label{eq:Chat_j_general}
\end{align}
and, substituting into the Laplace-transformed Eq.~(\ref{eq:Sd}),
\begin{eqnarray}
\hat S(s) = \frac{S(0)+\sum_j k_{\rm TF}|C_{\rm s}^{(j)}(B)|^2 C^{(j)}(0)/D_j(s)}{s+k_{\rm SF}+k_{\rm rad}-\sum_j k_{\rm SF}k_{\rm TF}|C_{\rm s}^{(j)}(B)|^4/D_j(s)},
\label{eq:Shat_general}
\end{eqnarray}
which reduces to Eq.~(\ref{eq:LaplaceS}) for $S(0)\neq0$ and $C^{(j)}(0)=0$. For $S(0)=0$ and $C^{(j)}(0)=|C_{\rm s}^{(j)}(B)|^2$, Eq.~(\ref{eq:Shat_general}) simplifies to
\begin{eqnarray}
\hat S(s) = \frac{k_{\rm TF}\sum_j|C_{\rm s}^{(j)}(B)|^4/D_j(s)}{s+k_{\rm SF}+k_{\rm rad}-k_{\rm SF}\sum_j k_{\rm TF}|C_{\rm s}^{(j)}(B)|^4/D_j(s)}.
\label{eq:Shat_reverse}
\end{eqnarray}

\begin{figure}
\centerline{
\includegraphics[width=1\columnwidth]{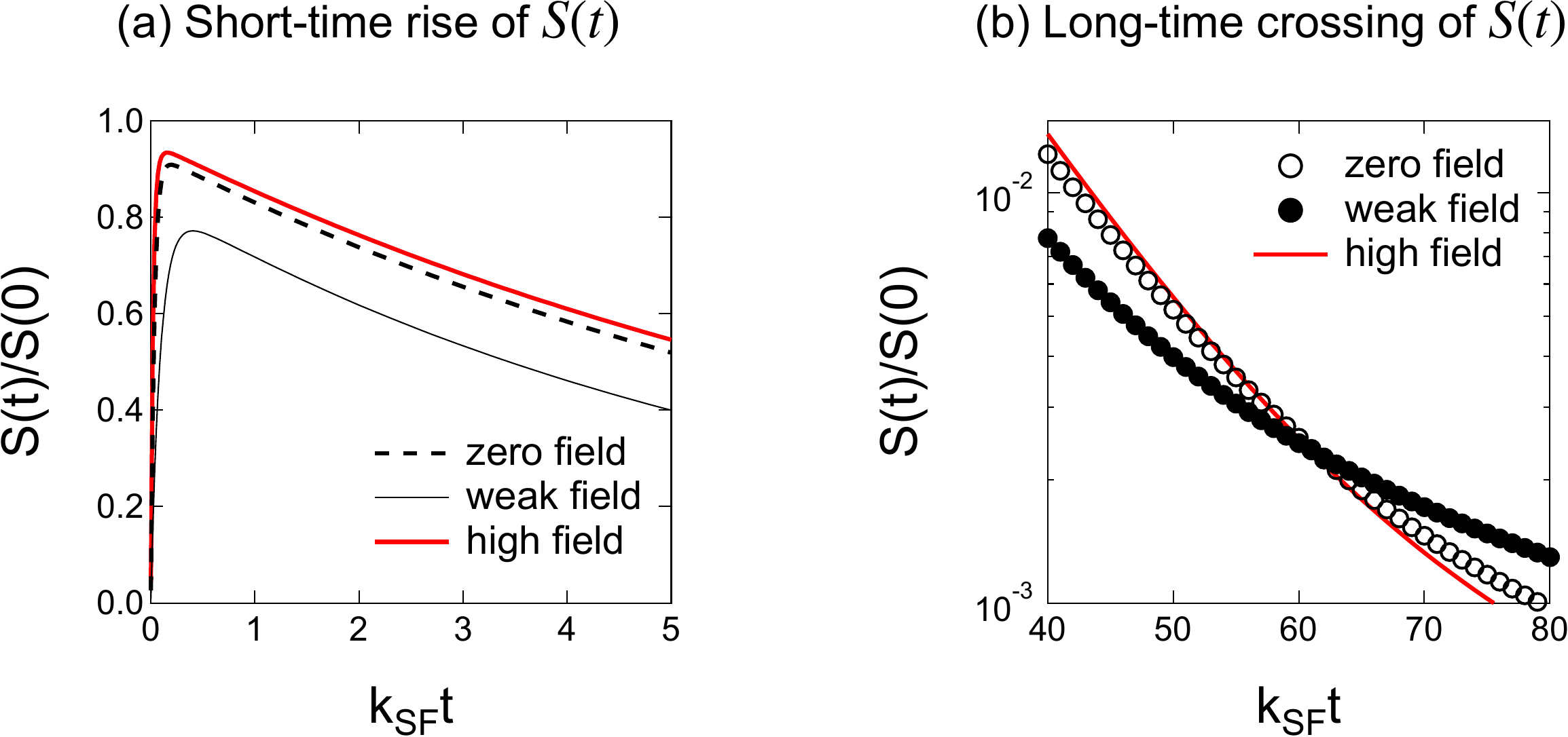}
}
\caption{ Relaxation trajectories $S(t)$ obtained from the field-matched triplet-pair initial condition $S(0)=0$, $C^{(j)}(0)=|C_{\rm s}^{(j)}(B)|^2$ [Eq.~(\ref{eq:Shat_reverse})], using the same parameters as Fig.~\ref{fig:Mpemba_cross}. (a) Short-time regime: $S(t)$ rises from $S(0)=0$ to a maximum generated by triplet fusion of the initially populated triplet pairs, whose height and timing follow the fusion efficiency of each field's singlet character (high field: highest, earliest peak; weak field: lowest, latest peak). (b) Long-time regime, same window as Fig.~\ref{fig:Mpemba_cross}: open circles, zero field; filled circles, weak field; red line, high field. The amplitude here is instead governed by dissociation of the triplet pairs in competition with fusion, followed by diffusion-controlled geminate recombination; trajectory crossing occurs between all three fields, near the same location as in Fig.~\ref{fig:Mpemba_cross}.}
\label{fig:reverse_ic}
\end{figure}

Numerically inverting Eq.~(\ref{eq:Shat_reverse}) by the Stehfest algorithm, using the same parameters as Fig.~\ref{fig:Mpemba_cross} and the same equal-character approximations of Secs.~\ref{sec:criterion_field} and~\ref{sec:criterion_highfield} for the zero-, weak-, and high-field cases, we find that, since $S(0)=0$ for every field, $S(t)$ first rises to a maximum before decaying [Fig.~\ref{fig:reverse_ic}(a)]. This initial rise is generated directly by triplet fusion of the initially populated triplet pairs: the high-field trajectory, whose concentrated singlet character fuses most efficiently, reaches the highest and earliest peak, while the weak-field trajectory, whose singlet character is spread thinly over all nine states, fuses least efficiently and reaches the lowest and latest peak -- the opposite ordering from that found at long times. At long times [Fig.~\ref{fig:reverse_ic}(b)], the amplitude of $S(t)$ is instead governed by dissociation of the triplet pairs occurring in competition with triplet fusion: pairs that dissociate before fusing subsequently re-encounter via diffusion and regenerate $S(t)$ through the same diffusion-controlled geminate recombination process discussed in Sec.~\ref{sec:modecomp}, producing the long-time power-law tail whose field-dependent amplitude reverses the short-time ordering and gives rise to trajectory crossing between all three fields, with the high-field trajectory crossing the zero-field trajectory between $k_{\rm SF}t=58$ and $k_{\rm SF}t=60$, and the weak-field trajectory crossing the zero-field trajectory between $k_{\rm SF}t=60$ and $k_{\rm SF}t=62$ -- essentially the same location as the crossing in Fig.~\ref{fig:Mpemba_cross}. This close agreement between the two calculations validates the time-scale-separation assumption used throughout the main text: treating the post-mixing triplet-pair state as an effective initial condition for the fusion-and-diffusion-controlled-recombination dynamics reproduces the same crossing obtained by evolving the full dynamics from $S(0)$, confirming that singlet fission followed by fast spin mixing is a valid experimental means of setting the (effectively field-dependent) initial condition for the process studied in this work.

\section*{Data availability}
No new experimental data were generated in this study. The numerical results can be reproduced from the equations and parameter values given in the manuscript.

\section*{Declaration of competing interest}
The author declares no competing financial interests or personal relationships that could have appeared to influence the work reported in this paper.

\section*{Funding}
This research did not receive any specific grant from funding agencies in the public, commercial, or not-for-profit sectors.

\section*{Declaration of generative AI and AI-assisted technologies}
During the preparation of this work, the author used ChatGPT and  Claude for language editing and improving readability. After using this tool, the author reviewed and edited the content as needed and takes full responsibility for the content of the publication.

\bibliographystyle{elsarticle-num} 
\bibliography{Mpemba}

\begin{thebibliography}{10}
\expandafter\ifx\csname url\endcsname\relax
  \def\url#1{\texttt{#1}}\fi
\expandafter\ifx\csname urlprefix\endcsname\relax\def\urlprefix{URL }\fi
\expandafter\ifx\csname href\endcsname\relax
  \def\href#1#2{#2} \def\path#1{#1}\fi

\bibitem{mpemba1969}
E.~B. Mpemba, D.~G. Osborne, Cool?, Phys. Educ. 4~(3) (1969) 172.
\newblock \href {http://dx.doi.org/10.1088/0031-9120/4/3/312}
  {\path{doi:10.1088/0031-9120/4/3/312}}.

\bibitem{lu2017mpemba}
Z.~Lu, O.~Raz, Nonequilibrium thermodynamics of the markovian mpemba effect and
  its inverse, Proc. Natl. Acad. Sci. U. S. A. 114~(20) (2017) 5083--5088.
\newblock \href {http://dx.doi.org/10.1073/pnas.1701264114}
  {\path{doi:10.1073/pnas.1701264114}}.

\bibitem{TEZA_26}
G.~Teza, J.~Bechhoefer, A.~Lasanta, O.~Raz, M.~Vucelja, Speedups in
  nonequilibrium thermal relaxation: Mpemba and related effects, Phys. Rep.
  1164 (2026) 1--97.
\newblock \href
  {http://dx.doi.org/https://doi.org/10.1016/j.physrep.2025.10.009}
  {\path{doi:https://doi.org/10.1016/j.physrep.2025.10.009}}.

\bibitem{Wang_26}
X.~Wang, J.~Su, J.~Wang,
  \href{https://link.aps.org/doi/10.1103/gjc2-8fkb}{Thermalization and
  mpemba-like patterns in effective temperature dynamics of strongly coupled
  dissipative quantum chaotic systems}, Phys. Rev. B 113 (2026) 045119.
\newblock \href {http://dx.doi.org/10.1103/gjc2-8fkb}
  {\path{doi:10.1103/gjc2-8fkb}}.
\newline\urlprefix\url{https://link.aps.org/doi/10.1103/gjc2-8fkb}

\bibitem{hayakawa2023}
A.~K. Chatterjee, S.~Takada, H.~Hayakawa, Quantum mpemba effect in a quantum
  dot with reservoirs, Phys. Rev. Lett. 131 (2023) 080402.
\newblock \href {http://dx.doi.org/10.1103/PhysRevLett.131.080402}
  {\path{doi:10.1103/PhysRevLett.131.080402}}.

\bibitem{hayakawa2024multiple}
A.~K. Chatterjee, S.~Takada, H.~Hayakawa, Multiple quantum mpemba effect:
  Exceptional points and oscillations, Phys. Rev. A 110 (2024) 022213.
\newblock \href {http://dx.doi.org/10.1103/PhysRevA.110.022213}
  {\path{doi:10.1103/PhysRevA.110.022213}}.

\bibitem{hayakawa2026}
H.~Hayakawa, S.~Takada, Mpemba effect in a two-dimensional bistable potential
  (2026).
\newblock \href {http://arxiv.org/abs/2603.24148} {\path{arXiv:2603.24148}}.

\bibitem{brownian_singlewell2023}
A.~Biswas, R.~Rajesh, Mpemba effect for a brownian particle trapped in a single
  well potential, Phys. Rev. E 108 (2023) 024131.
\newblock \href {http://dx.doi.org/10.1103/PhysRevE.108.024131}
  {\path{doi:10.1103/PhysRevE.108.024131}}.

\bibitem{langevin_metastability2023}
A.~Biswas, R.~Rajesh, A.~Pal, Mpemba effect in a langevin system: Population
  statistics, metastability, and other exact results, J. Chem. Phys. 159 (2023)
  044120.

\bibitem{active_markov2025}
A.~Biswas, A.~Pal, Mpemba effect on non-equilibrium active markov chains, Phys.
  Rev. E 111 (2025) 054136.
\newblock \href {http://arxiv.org/abs/2403.17547} {\path{arXiv:2403.17547}}.

\bibitem{active_brownian_notrap2025}
A.~Biswas, R.~Rajesh, Mpemba effect in the relaxation of an active brownian
  particle in a trap without metastable states, J. Chem. Phys. 162~(3) (2025)
  034115.
\newblock \href {http://dx.doi.org/10.1063/5.0246857}
  {\path{doi:10.1063/5.0246857}}.

\bibitem{granular_maxwell2020}
A.~Biswas, V.~V. Prasad, O.~Raz, R.~Rajesh, Mpemba effect in driven granular
  maxwell gases, Phys. Rev. E 102~(1) (2020) 012906.
\newblock \href {http://dx.doi.org/10.1103/PhysRevE.102.012906}
  {\path{doi:10.1103/PhysRevE.102.012906}}.

\bibitem{granular_anisotropic2021}
A.~Biswas, V.~V. Prasad, R.~Rajesh, Mpemba effect in an anisotropically driven
  granular gas, EPL 136~(4) (2021) 46001.
\newblock \href {http://dx.doi.org/10.1209/0295-5075/ac2d54}
  {\path{doi:10.1209/0295-5075/ac2d54}}.

\bibitem{trapped_ion_inverse2024}
S.~A. Shapira, Y.~Shapira, J.~Markov, G.~Teza, N.~Akerman, O.~Raz, R.~Ozeri,
  Inverse mpemba effect demonstrated on a single trapped ion qubit, Phys. Rev.
  Lett. 133 (2024) 010403.
\newblock \href {http://dx.doi.org/10.1103/PhysRevLett.133.010403}
  {\path{doi:10.1103/PhysRevLett.133.010403}}.

\bibitem{burdett2013}
J.~J. Burdett, G.~B. Piland, C.~J. Bardeen, Magnetic field effects and the role
  of spin states in singlet fission, Chem. Phys. Lett. 585 (2013) 1--10.
\newblock \href
  {http://dx.doi.org/https://doi.org/10.1016/j.cplett.2013.08.036}
  {\path{doi:https://doi.org/10.1016/j.cplett.2013.08.036}}.

\bibitem{piland2013}
G.~B. Piland, J.~J. Burdett, D.~Kurunthu, C.~J. Bardeen, Magnetic field effects
  on singlet fission and fluorescence decay dynamics in amorphous rubrene, J.
  Phys. Chem. C 117 (2013) 1224.

\bibitem{seki2018}
K.~Seki, Y.~Sonoda, R.~Katoh, Diffusion-mediated delayed fluorescence by
  singlet fission and geminate fusion of correlated triplets, J. Phys. Chem. C
  122~(22) (2018) 11659--11670.
\newblock \href {http://dx.doi.org/10.1021/acs.jpcc.8b02234}
  {\path{doi:10.1021/acs.jpcc.8b02234}}.

\bibitem{shushin2018}
A.~Shushin, Kinetic curves crossing effect. manifestation of the effect in
  singlet fission in organic semiconductors, Chem. Phys. Lett. 712 (2018) 165
  -- 170.
\newblock \href
  {http://dx.doi.org/https://doi.org/10.1016/j.cplett.2018.09.071}
  {\path{doi:https://doi.org/10.1016/j.cplett.2018.09.071}}.

\bibitem{seki2021}
K.~Seki, T.~Yoshida, T.~Yago, M.~Wakasa, R.~Katoh, Geminate delayed
  fluorescence by anisotropic diffusion-mediated reversible singlet fission and
  triplet fusion, J. Phys. Chem. C 125~(6) (2021) 3295--3304.
\newblock \href {http://dx.doi.org/10.1021/acs.jpcc.0c10582}
  {\path{doi:10.1021/acs.jpcc.0c10582}}.

\bibitem{johnson1970}
R.~C. Johnson, R.~E. Merrifield, Effects of magnetic fields on the mutual
  annihilation of triplet excitons in anthracene crystals, Phys. Rev. B 1
  (1970) 896.

\bibitem{merrifield1971}
R.~E. Merrifield, Magnetic effects on triplet exciton interactions, Pure Appl.
  Chem. 27 (1971) 481.

\bibitem{steiner1989}
U.~E. Steiner, T.~Ulrich, Magnetic field effects in chemical kinetics and
  related phenomena, Chem. Rev. 89 (1989) 51.

\bibitem{nonMarkovian2025}
D.~J. Strachan, A.~Purkayastha, S.~R. Clark, Non-markovian quantum mpemba
  effect, Phys. Rev. Lett. 134 (2025) 220403.
\newblock \href {http://dx.doi.org/10.1103/PhysRevLett.134.220403}
  {\path{doi:10.1103/PhysRevLett.134.220403}}.

\bibitem{Bechhoefer2021}
J.~Bechhoefer, A.~Kumar, R.~Ch{\'e}trite, A fresh understanding of the mpemba
  effect, Nat. Rev. Phys. 3~(8) (2021) 534--535.
\newblock \href {http://dx.doi.org/10.1038/s42254-021-00349-8}
  {\path{doi:10.1038/s42254-021-00349-8}}.

\bibitem{Ares2025}
F.~Ares, P.~Calabrese, S.~Murciano, The quantum mpemba effects, Nat. Rev. Phys.
  7~(8) (2025) 451--460.
\newblock \href {http://dx.doi.org/10.1038/s42254-025-00838-0}
  {\path{doi:10.1038/s42254-025-00838-0}}.

\bibitem{redner_2001}
S.~Redner, A Guide to First-Passage Processes, Cambridge University Press,
  2001.
\newblock \href {http://dx.doi.org/10.1017/CBO9780511606014}
  {\path{doi:10.1017/CBO9780511606014}}.

\bibitem{Abramowitz}
M.~Abramowitz, I.~A. Stegun, {H}andbook of {M}athematical {F}unctions, Dover,
  New York, 1972.

\bibitem{Feller_71}
W.~Feller, An Introduction to Probability Theory and Its Applications, Vol. 2,
  2nd Edition, Wiley, New York, NY, 1971.

\bibitem{merrifield1968}
R.~E. Merrifield, Theory of magnetic field effects on the mutual annihilation
  of triplet excitons, J. Chem. Phys. 48 (1968) 4318.

\bibitem{Stehfest1970_47}
H.~Stehfest, Algorithm 368: Numerical inversion of laplace transforms [d5],
  Commun. ACM 13~(1) (1970) 47--49.

\bibitem{Stehfest1970_624}
H.~Stehfest, Remark on algorithm 368: Numerical inversion of laplace
  transforms, Commun. ACM 13~(10) (1970) 624--624.

\bibitem{pontus2025}
A.~Nava, R.~Egger, Pontus-mpemba effects, Phys. Rev. Lett. 135 (2025) 140404.
\newblock \href {http://dx.doi.org/10.1103/hhgj-89gj}
  {\path{doi:10.1103/hhgj-89gj}}.

\bibitem{descartes2026}
A.~Santos, The mpemba effect in the {D}escartes protocol: a time-delayed
  {N}ewton's law of cooling approach, J. Phys. A: Math. Theor. 59 (2026)
  145201.
\newblock \href {http://dx.doi.org/10.1088/1751-8121/ae57ed}
  {\path{doi:10.1088/1751-8121/ae57ed}}.

\bibitem{distance_measure_granular2023}
A.~Biswas, V.~V. Prasad, R.~Rajesh, Mpemba effect in driven granular gases:
  Role of distance measures, Phys. Rev. E 108 (2023) 024902.
\newblock \href {http://dx.doi.org/10.1103/PhysRevE.108.024902}
  {\path{doi:10.1103/PhysRevE.108.024902}}.

\bibitem{thermomajorization2025}
T.~Van~Vu, H.~Hayakawa, Thermomajorization mpemba effect, Phys. Rev. Lett. 134
  (2025) 107101.
\newblock \href {http://dx.doi.org/10.1103/PhysRevLett.134.107101}
  {\path{doi:10.1103/PhysRevLett.134.107101}}.

\bibitem{yago_16}
T.~Yago, K.~Ishikawa, R.~Katoh, M.~Wakasa, Magnetic field effects on triplet
  pair generated by singlet fission in an organic crystal: Application of
  radical pair model to triplet pair, J. Phys. Chem. C 120 (2016) 27858--27870.
\newblock \href {http://dx.doi.org/10.1021/acs.jpcc.6b09570}
  {\path{doi:10.1021/acs.jpcc.6b09570}}.

\end{thebibliography}






\end{document}